%% file: Tubiana_et_al_Synonymous_Mutations_Erase_Viral_RNA_Compactness_v1_arXiv.tex
\documentclass[pre,onecolumn,aps,floats,floatfix,unsortedaddress,superscriptaddress]{revtex4-1}
\usepackage{natbib}
\usepackage{graphicx}
\usepackage{color}
\usepackage{amsfonts}
\usepackage{amssymb,amsmath}
\usepackage{mathrsfs}
\usepackage[english]{babel}
\usepackage{babelbib}
\usepackage{booktabs} %professional-like tables
\usepackage{longtable}

\usepackage{xr}
\newcommand{\MLD}{\langle\mathrm{MLD}\rangle}

\usepackage[minuso]{stmaryrd}
\usepackage[T1]{fontenc}

\begin{document}

\title{Synonymous mutations reduce genome compactness in icosahedral ssRNA viruses}
\author{Luca Tubiana}
\email{luca.tubiana@ijs.si}
\thanks{Corresponding author}
\affiliation{Department of Theoretical Physics, Jo\v zef Stefan Institute, SI-1000 Ljubljana, Slovenia}
\author{An\v{z}e Lo\v{s}dorfer Bo\v{z}i\v{c}}
\affiliation{Department of Theoretical Physics, Jo\v zef Stefan Institute, SI-1000 Ljubljana, Slovenia}
\affiliation{Max Planck Institute for Biology of Ageing, Joseph-Stelzmann-Str.\ 9b, D-50931 Cologne, Germany}
\author{Cristian Micheletti}
\affiliation{SISSA, Via Bonomea 265, I-34136 Trieste, Italy}
\author{Rudolf Podgornik}
\affiliation{Department of Theoretical Physics, Jo\v zef Stefan Institute, SI-1000 Ljubljana, Slovenia}
\affiliation{Department of Physics, Faculty of Mathematics and Physics, University of Ljubljana, SI-1000 Ljubljana, Slovenia}
\affiliation{Department of Physics, University of Massachusetts, Amherst, MA 01003}
\date{\today}

\begin{abstract}
Recent studies have shown that single-stranded viral RNAs fold into more
compact structures than random RNA sequences with similar chemical
composition and identical length. Based on this comparison it has been
suggested that wild-type viral RNA may have evolved to be atypically
compact so as to aid its encapsidation and assist the viral assembly
process. In order to further explore the compactness selection hypothesis,
we systematically compare the predicted sizes of more than one hundred wild-type
viral sequences with those of their mutants, which are evolved {\em in
silico} and subject to a number of known evolutionary constraints. In
particular, we enforce mutation synonynimity, preserve the codon-bias, and
leave untranslated regions intact. It is found that progressive
accumulation of these restricted mutations still suffices to completely
erase the characteristic compactness imprint of the viral RNA genomes,
making them in this respect physically indistinguishable from randomly
shuffled RNAs. This shows that maintaining the physical compactness of
the genome is indeed a primary factor among ssRNA viruses evolutionary
constraints, contributing also to the evidence that synonymous
mutations in viral ssRNA genomes are not strictly neutral.
\end{abstract}

\maketitle

\section*{INTRODUCTION}

Minimalistic organisms, such as single-stranded (ss)RNA viruses, are ideally suited to
investigate how the three-dimensional organization of the genome -- and
not only its sequence composition -- is subject to selective evolutionary
pressure. We recall, for instance, that several structural features are robustly
maintained in the highly-mutating ssRNA viruses. These include RNA structures
acting as signals for translation~\cite{olsthoorn1996}, for transcription initiation~\cite{klovins1998},
or as packaging signals to initiate the self-assembly of the
virion~\cite{Dykeman2013,Dykeman2014}. Other conserved structures have also been
identified~\cite{Simmonds2004,Sanjuan2011,Cuevas2012}, including long-range
interactions between different genomic regions of
RNA~\cite{Simmonds2004,Davis2008}, whose role in the virus life cycle is
still unknown.

The preservation of these structural features must act as a powerful
constraint on viable RNAs, together with the multiple other, often competing, selection pressures~\cite{Holmes2009,Belshaw2007,Eigen2000}.
The evolutionary mechanisms which maintain the viral protein phenotype
clearly impact the genome chemical composition more directly, by
largely restricting those mutations which have a deleterious effect on the 
encoded proteins~\cite{Wylie:2011:PNAS,Chen:2009:Genetics,Duffy2008,Gong2013}. 
On the other hand, synonymous mutations, i.e., mutations that do not change the amino acid sequences 
encoded by the genes, are neutral with regard to these mechanisms, but still have an impact on the structural features of RNAs.

It is increasingly becoming recognised that the mechanisms which may constrain
synonymous mutations extend beyond the aforementioned conservation of specific genome structures, and are
underpinned by general physico-chemical constraints. The latter mostly stem from the polymeric nature of the gene-carrying
macromolecules and their steric and electrostatic self-interactions, as
well as interactions with the capsid proteins~\cite{Hyeon2006,Marenduzzo2009,Marenduzzo22112013, Tandogan2014}. These molecular
interactions can be long-ranged and depend crucially on the pH of the
local aqueous solution environment~\cite{Nap2014}, conferring virions the ability to
assemble and disassemble spontaneously at proper bathing solution
conditions~\cite{Caspar1990,Bruinsma2003,Reguera2004,Singh2003,Nguyen2007,Castellanos2012,Roos2012,Polles2013},
and the ability to recognize and selectively encapsidate only viral RNA
even in the absence of packaging
signals~\cite{Tandogan2014,Cadena2012,ComasGarcia2012,Perlmutter2013,Harvey2013}.

In this study we focus on a general and major structure-related
selection constraint, namely the feasibility to efficiently package
viral RNA inside the capsid, and address its competition with
sequence-based selection mechanisms. The overarching question is whether the viral RNA sequence has evolved
not only for encoding a specific protein phenotype but also for
promoting an innate fold of the free (unencapsidated) viral RNA itself
that is primed for efficient encapsidation.

Major advances towards solving this important conundrum have been
recently made by comparing the predicted equilibrium properties of ssRNA
folds of several icosahedral viruses with those of random RNA sequences
with similar length and nucleotide composition. By using general
arguments based on the scaling properties of linear~\cite{Yoffe2008}
and/or branched polymers~\cite{Fang2011}, the folded wild-type (WT)
viral RNA was shown to be significantly more compact than random
nucleotide sequences.  In addition -- and most notably -- the average
radius of gyration of WT RNA genomes was found to exceed only slightly
the inner radius of the fully-assembled capsid~\cite{Gopal2012}.

In this context, a key and still open problem relates to the extent to
which the selective pressure for easily encapsidable RNA genomes
directly competes with the other sequence-based mechanisms that are simultaneously at
play for selecting biologically-viable viral RNA.  As a matter of fact,
the enhanced compactness of viral RNA has so far been established only
by comparison against random sequences that do not retain any specific
viral-like characteristics except from the overall nucleotide
composition. As the volume of the sequence ``phase space'' that is
accessible to viable viral RNA sequences is actually vanishingly small
compared to the available combinatorial phase space of random
sequences, it is crucial to ascertain the implications of introducing
realistic sequence constraints into the picture. Such
constraints could even affect the properties of the associated folds to
the point of implying genome compactness, which would make the assumption
of a distinct selection principle based on RNA compactness superfluous.

To address these issues we consider the implications of
constrained mutations that conserve the encoded protein phenotype and
the viral-like nucleotide composition on the compactness of viral RNA genomes. This allows us to examine the
concurrence, or possibly the incompatibility, of sequence- and
structure-based {\em parallel selection mechanisms}, and to ascertain
whether the conservation of RNA compactness is among the causes of the
sensitivity of ssRNA viruses to synonymous mutations.

Specifically, we consider 128 viral RNA sequences and evolve them %4 satellites are not considered
synthetically by accumulating exclusively {\em synonymous} point-wise
mutations, measuring their impact on the properly quantified compactness
of the genome. We recall that the constraint of synonymity,
i.e., considering only codons that encode for the same amino acids, is
particularly severe for viral RNA because of both the high gene density
and the frequent presence of overlapping reading frames.

Our study unequivocally shows that, at least for the viruses studied,
the accumulation of strictly synonymous mutations -- even if they are
sparse -- is sufficient to cause a systematic drift of the properly
quantified compactness of the genome towards values comparable to those
of unrestricted random sequences that are systematically much larger
than those of the WT genomes. By focusing on the
mutational dynamics of four viral genomes we show that while mutating as
few as 5~\% of a genome is enough to erase its compactness, there
is still a non-negligible portion of the sequence space in the vicinity
of the WT sequence in which the genomes are at least as compact as the
WT genome, while still coding for the correct proteins.

Furthermore, we show that the typical WT RNA compactness is related neither
to the codon usage biases present in viral genomes nor to the particular
sequences of the untranslated regions (UTRs) present at the 5' and 3' ends
of the genomes.
These results provide {\em a posteriori} evidence that the same viral RNA sequence can encode not only for the expression of the proper protein complement, exposed to canonical selection pressure mechanisms, but can on another level also prime the optimal physico-chemical genome-packing organization.

\iffalse The paper is structured as follows: First, in
Sec.~\ref{ssec:meth:fold} we present the basic methods employed for
producing the random RNA sequences, and folding both the random and the
viral ssRNA sequences. In Sec.~\ref{ssec:meth:WT} we briefly present the
dataset we use, and in Sec.~\ref{ssec:meth:mut} we describe the simple
models of synonymous point mutations whose importance we assess in
relation to their effect on the genome compactness. Then in
Sec.~\ref{sec:results} we compare the results for the viral Maximum
Ladder Distances with the ones obtained by Yoffe et
al.~\cite{Yoffe2008}, and present the effect of point mutations on the
compactness of the secondary structures of viral RNAs. We conclude with
a discussion and implications of our results in
Sec.~\ref{sec:disc}. \RP{If necessary this paragraph could go out.}  \fi

%\begin{materials}
\section*{MATERIALS AND METHODS}

\subsection*{Wild-type viral sequences}\label{ssec:meth:WT}
Viral ssRNA sequences were obtained from the NCBI nucleotide database~\cite{nucleotide}. The dataset we use includes positive-strand ssRNA viruses from the following families: Tymoviridae (from the order Tymovirales), Flaviviridae, Caliciviridae, Picornaviridae, Comovirinae, Bromoviridae, and Tombusviridae~\cite{VIRALZONE}. All the viruses considered have icosahedral capsids, the majority of them with triangulation number $T=3$. Most of the families in the dataset have monopartite genome, with the exception of Comovirinae, which have a bipartite genome, and Bromoviridae, which have a tripartite genome~\cite{VIRALZONE}. Comovirinae pack the two segments, denoted RNA1 and RNA2, into separate virions; the two largest RNA segments of Bromoviridae genome, denoted RNA1 and RNA2, are also packed into separate virions, and we thus consider only these two segments. All the considered viruses use the eukaryotic genetic code and their genes
have no reading gaps. Several sequences among those we consider
also have overlapping reading frames, which are known to impose further
evolutionary constraints increasing the deleterious effects of
mutations~\cite{SimonLoriere2013,Chirico2010}. With these restrictions taken into account, the final dataset of analyzed
sequences contains 128 viral genomes (compiled in Supporting Information (SI) Table S1).

\subsection*{Synonymous point mutations}\label{ssec:meth:mut}
Extended models of sequence evolution of overlapping genes can account for the codependency of the nucleotide substitution process in two reading frames~\cite{Pedersen2001,Chung2007}, but are based in computationally very intense simulations and are not always applicable to large sequence datasets. Since in the present study we are interested in the statistical properties across various viral families, we adopt a much simpler model which simply conserves the produced amino acids in all reading frames. 

Mutated viral ssRNA sequences are obtained using a Monte Carlo (MC) scheme designed to simulate synonymous point substitutions while also conserving dinucleotide frequencies. Starting from a WT sequence, a point substitution is introduced at every step and accepted or rejected using a Metropolis algorithm. Substitutions which change the amino acids encoded by the genes and are thus non-synonymous are rejected. To preserve the dinucleotide frequencies we additionally introduce a fictitious energy related to the viral dinucleotide odd-ratios~\cite{Nussinov:1981:JBioChem}:
\begin{equation}
\label{eq:kxy}
  E =  \sum_{XY} K_{XY}\left[O(XY) - O_{WT}(XY)\right]^2,
\end{equation}
where
\begin{equation}
O(XY) =  \frac{N(XY)}{N(X)N(Y)}N,\quad X,Y\in\lbrace A,U,G,C \rbrace.
\end{equation}
Here, $N(XY)$ is the number of $XY$ pairs, $N(X),N(Y)$ are the numbers of $X$ and $Y$ nucleotides in the sequence, and $N$ is the total length of the RNA sequence.

The values of the constants $K_{XY}$ are chosen in such a way that a
considerable portion (but not all) of the proposed sequences have
dinucleotide odd-ratios lying within $1.5\Delta Q$, where $\Delta Q$ is
the interquartile distance evaluated on the $O_{WT}(XY)$ distribution of
the corresponding viral family (see SI for additional information). We produce
an extensive ensemble of point mutations ($\sim 10^{9}$) to
ensure an appropriate sampling of the sequence space. Sequences are
sampled every $100 N$ mutations to ensure they are uncorrelated, and
filtered a posteriori to have all odd-ratios within $1.5\Delta Q$. For
every WT viral sequence we generate a set of 500 to 2000 mutated
sequences and finally characterise the spatial compactness of the
associated fold by computing the thermally averaged maximum ladder
distance, $\MLD$, described in a later subsection.

As an additional check we also produce synonymous substitutions using the Fisher-Yates shuffling algorithm~\cite{Durstenfeld:1964:ACM,Knuth-AoCP2} -- in this way, the exact chemical composition of the sequences is conserved, although the dinucleotide odd-ratios are not. %; these results are presented in the SI Fig. S3.
While much more complex models for the nucleotide substitutions exist (see for instance the review by Anisimova and Kosiol~\cite{Anisimova2009} and references therein), we chose these simple ones that conserve the chemical composition of the sequences as they are sufficient to prove our point, and can most importantly be applied in the same manner to all the genomes we considered.
%\subsection*{Mutation dynamics}
% XXX Maybe it's better to avoid the ``mutation dynamics'' heading because the 
% next heading is again related to mutations. It's simpler to incorporate it here.

To investigate the effect of progressively accumulating mutations on viral RNA compactnes, quantifyed by the MLD, we first choose the $K_{XY}$ values in such a way that all produced sequences obey the dinucleotide constraints. The generated MC trajectories are then sampled every $N/100$ steps. This sampling produces strongly correlated sequences which show the evolution of the genome MLDs toward the values of their random counterparts.

%with the following constraints: i) the mutated sequences are the viral
%genomes themselves, ii) every exchange of two nucleotides is accepted only
%if it does not affect the amino-acid sequences coded by the genome. \LT{how
%many folds? how many mutations? how does the dynamic work?}. Our
%substitution model is indeed very simple, but enough to enforce  fixed
%chemical composition of the sequences. 

\subsection*{Synonymous mutations preserving codon bias} \label{ssec:meth:cbs}
As an optional additional constraint, we fix the WT codon population by shuffling equivalent codons, as 
done in Ref.~\cite{Gu2010}. The shuffling is performed at the gene-wise level by first enumerating and pooling the synonymous 
codons  in the WT gene sequence. Each codon in the latter is then replaced by one picked randomly from 
its synonymous pool. The pools are thus progressively depleted until all reassignments are completed, 
as in the standard Fisher-Yates shuffling algorithm~\cite{Durstenfeld:1964:ACM,Knuth-AoCP2}.  This shuffling procedure, which clearly 
preserves the WT codon bias at the gene level is applicable to viral genomes without overlapping genes, which are 86 in our case.

\subsection*{Random RNA sequences} \label{ssec:meth:ran}
Random ssRNA sequences, used to obtain the scaling law for the MLD of
random RNAs, are produced by shuffling RNA sequences with the
Fisher-Yates algorithm~\cite{Durstenfeld:1964:ACM,Knuth-AoCP2}.  Random
numbers, here as well as in the rest of the paper, are generated by the
SIMD-oriented Fast Mersenne Twister (SFMT) random generator, version
1.4~\cite{SFMT}. The SFMT has a period of $2^{216091}-1$, which suffices
to produce random permutations of even 10 knt long RNA sequences. We use
the same viral-like composition for the random sequences as in
Ref.~\cite{Yoffe2008}, that is, $0.26$ A, $0.28$ U, $0.24$ G, and $0.22$
C, to obtain the scaling law for random viral-like RNAs. This average
composition is computed excluding Tymoviridae, which differ
significantly in their composition. For the Tymoviridae family, we use
the averaged composition of the viruses in our sample belonging to this
family only (see SI Table S1 for the list), with the corresponding
nucleotide composition: $0.219$ A, $0.254$ U, $0.163$ G, $0.364$ C.

\subsection*{Maximum Ladder Distance (MLD)}\label{ssec:meth:MLD}

In order to investigate the possibility that synonymous substitutions, while being neutral with respect to the encoded protein complement, can affect the secondary structure of viral RNA, we use  the (thermally averaged) MLD, a quantitative, albeit coarse-grained indicator of the compactness of RNA folds introduced by Yoffe and coworkers~\cite{Yoffe2008}. While the MLD of random RNAs with viral-like nucleotide composition follows a simple scaling law, the MLDs of viral ssRNA genomes are on the other hand significantly lower, indicating that their folds are more compact than those of random RNAs. 

\begin{figure*}[t]
\begin{center}
\includegraphics[width=\textwidth]{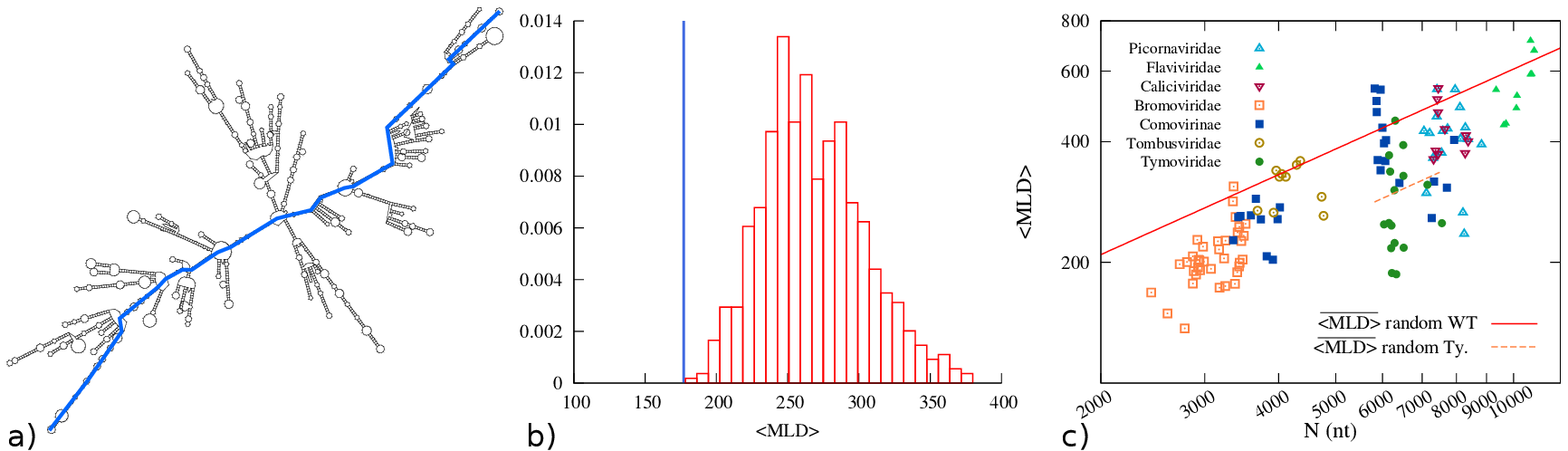}     %% color online
\end{center}
\caption{a) Example of a typical fold of the entire brome mosaic
  virus (BMV) RNA2 sequence. The maximum ladder distance (MLD) of the folded
  sequence is highlighted.  b) Thermally averaged MLD, $\MLD$, of the WT BMV RNA2 sequence (blue line) and the
  distribution of $\MLD$s obtained for random RNA sequences
  of same length and composition as the WT sequence.  c) $\MLD$ of viral ssRNA
  sequences versus the sequence length $N$ (in nucleotides).  Different
  virus families are represented by different colors and symbols. The
  red solid line shows the power law of Eq.~(\ref{eq:randMLD}) for the
  expected values of $\MLD$ for random RNA sequences, constrained only by
  their overall viral-like nucleotide composition. Due to their atypical
  nucleotide composition, Tymoviridae are not represented by
  Eq.~(\ref{eq:randMLD}), and the corresponding scaling law for
  Tymoviridae-like random RNA sequences, $\overline{\MLD}_{{\mathrm Ty}}(N) = (0.92\pm0.44)\times N^{0.669\pm 0.054}$, is shown with
  an orange dashed line. See SI for further information. To see this figure
  in color, go online.
\label{fig:1}}
\end{figure*}

When treating the RNA as an ideal linear polymer, one can compute its MLD when mapped to an ideal graph~\cite{Yoffe2008,Bundschuh2002}: For every pair of nucleotides $i$
and $j$ in an RNA sequence we compute the ladder distance, i.e., the number of steps on the ladder which separates the two nucleotides on the folded RNA. The maximum value of all the ladder distances in a fold is then its MLD; an example is shown in Fig.~\ref{fig:1}(a). By treating the MLD contour as the backbone of a linear polymer chain, this provides a measure of compactness/extendedness of the RNA molecule, even though it is not a direct measure of the three-dimensional size of the RNA. This simple measure yields the same scaling relationships as in the case when one treats the RNA as an ideal branched polymer, computing its root-mean-square radius of gyration to determine its extendedness~\cite{Fang2011}.

The secondary structures of viral and random RNA sequences for which we
determine their MLDs are obtained by folding the sequences with the {\tt
RNAsubopt} program available in the ViennaRNA Package, version
2.1~\cite{VIENNARNA}. Due to the length of viral RNA, a population of
different folds having comparable energy is expected. Therefore, instead of
looking for the minimum energy fold, we produce $500$ folds at thermal 
equilibrium for every RNA sequence. This results in a thermal average for the MLD of every sequence, obtained by averaging over this ensemble.
%\end{materials}

\section*{RESULTS}\label{sec:results}

\subsection*{Validation: compactness of WT and random RNA sequences}

As a starting point for our analysis we considered an extensive set of
128 WT viral sequences listed in supporting information (SI) Table S1. We characterised their
compactness by following the method introduced by Yoffe et al.~\cite{Yoffe2008}, which entails two steps, detailed in the Materials and Methods section. The first step consists of computing an ensemble of several hundred representative
planar RNA folds using the ViennaRNA package~\cite{VIENNARNA}.  Next, one
calculates the maximum ladder distance (MLD) of each fold.  We recall that the
ladder distances are obtained by considering in turn all possible pairs of
nucleotides and identifying their shortest connecting path, i.e., the one
with the minimal number of ``rungs on the ladder'' along the duplexed
parts of the folds. The number of rungs of the longest minimal path is
the MLD, an example of which is shown in Fig.~\ref{fig:1}(a).

As discussed in Refs.~\cite{Yoffe2008} and~\cite{Fang2011}, the thermal
average of the MLD, denoted by $\MLD$, is a viable, albeit coarse-grained proxy for the
equilibrium spatial compactness of a folded sequence. Since it can be
calculated by highly efficient algorithms, it is particularly apt for
numerical implementation in extensive enumerative contexts such as the
present one.

The comparison of the $\MLD$s computed for the 128 viral sequences considered in our study with the $\MLD$s of random sequences with viral-like nucleotide
composition (see Materials and Methods) conforms to the earlier
conclusion of Yoffe et al.~\cite{Yoffe2008} that WT RNA genomes have an
enhanced fold compactness compared to arbitrary RNA sequences. This
point is illustrated in Figs.~\ref{fig:1}(b) and (c). As can be seen in
Fig~\ref{fig:1}(c), the $\MLD$s of random RNA sequences, additionally
averaged over several possible mutations, follow the power law
\begin{equation}
\label{eq:randMLD}
\overline{\MLD}(N)=(1.365\pm0.05)\times N^{0.662\pm0.004},
\end{equation}
\noindent where the overline indicates the additional averaging over different possible mutations.
On the other hand, the $\MLD$s of WT sequences are almost always more compact than the corresponding random
values given by Eq.~(\ref{eq:randMLD}). We also note that the parameters of the power law given by Eq.~(\ref{eq:randMLD})
are in good accord with the findings of Ref.~\cite{Yoffe2008}.

\subsection*{Compactness of WT and synonymously-mutated RNA sequences}

Since the fixation of mutations in viral genomes is subject to a number
of evolutionary pressures, the fact that WT RNA sequences of icosahedral
viruses tend to be more compact than predicted by Eq.~(\ref{eq:randMLD})
is not enough to conclude that they have been evolutionary selected for
optimal compactness. In fact, the sequence space
accessible to random mutations is unrealistically large because it does
not account for the several selection constraints that viable RNA
sequences have to obey.

Arguably, the most severe of such constraints reflects the necessity for the viruses to preserve their protein phenotype. Accordingly, we explore its implications for genome compactness
by considering only sequences which encode for the same proteins as the WT
RNA. This amounts to restricting our considerations only to
the rather limited combinatorial subspace of {\em
synonymous variants} of WT viral RNA sequences.

%From the observation that the MLD and with it the spatial extension of
%the WT RNAs is consistently smaller than the one belonging to the
%randomized sequences, one would be compelled to conclude that WT
%genomes are characterized by a certain size optimality. However, the randomised sequences analyzed above are required to respect only a viral-like nucleotide composition constraint. Thus, the phase
%space volume of these random sequences is unrealistically large
%compared to the one actually accessible to viral sequences, whose viable
%space of mutations is critically constrained by evolutionary mechanisms.

%In fact, viable viral sequences are subject to strong constraints 
%that impact the RNA nucleotide composition either directly
%(e.g., packaging, transcription or initiation signals, fold organization)
%or indirectly through the primary amino acid sequence of the encoded proteins. The latter constraint is particularly severe because the relatively
%short viral RNAs have a very high gene density. We explore its
%implications for genome compactness by considering only sequences which
%encode the same proteins as the WT RNA. This amounts to restricting our considerations only to
%the combinatorial subspace of {\em synonymous variants} of WT viral RNA sequences.

We recall that synonymous mutations originate in the degenerate mapping
of the 61 possible codons, which are nucleotide triplets, to the 20
canonical amino acids. Equivalent codons typically differ only at the
third nucleotide~\cite{Lehninger}.  Accordingly, we shall assume, for simplicity, that the
A, U, G, and C nucleotides can appear with equal probability at the
third codon position, one can estimate that two synonymous versions of a
gene have a nucleotide sequence identity of about 75~\%. Since, in the set of
viruses considered in our study, on average (90 $\pm$ 7)~\% of the
genome codes for at least one gene, and additionally assuming for
simplicity that the four nucleotides have equal probability in the
non-coding region which we are not constraining, we can estimate that at
least around 66-73~\% of the whole genome will be conserved under
synonymous mutation flow.

\begin{figure}[th]
\begin{center}
\includegraphics[width=0.5\columnwidth]{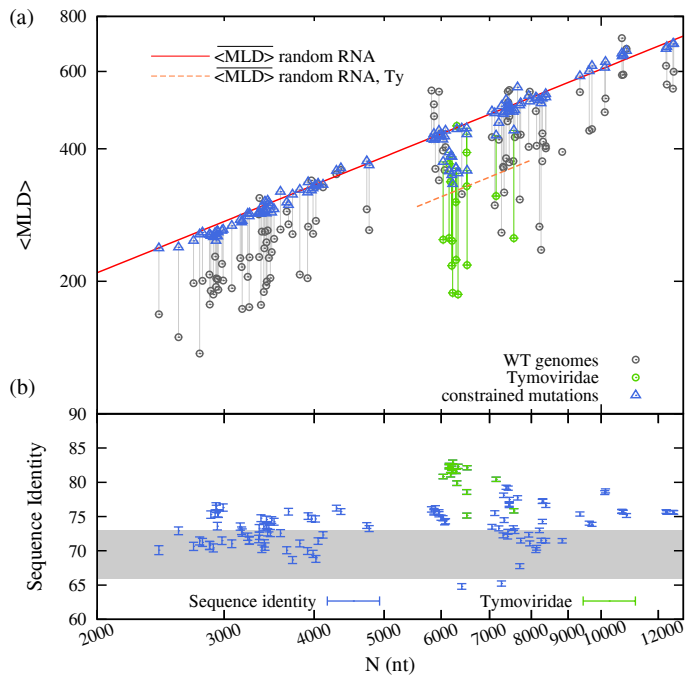}     %% color online
\end{center}
\caption{ a) Influence of synonymous point mutations on MLD. The $\MLD$s of
  WT viral sequences from Fig.~\ref{fig:1}(b) are shown as gray
  circles, and the $\overline{\MLD}$s of synonymously mutated sequences as blue triangles.
Scaling laws for $\overline{\MLD}$ of random RNA sequences with viral-like and Tymoviridae-like composition
  are shown as in Fig.~\ref{fig:1}. b) The average
  degree of sequence identity between the mutated and WT
  sequences. The gray shaded area indicates the values one would expect if only
  one in three nucleotides were allowed to mutate in the coding regions of
  the genomes.  Note that Tymoviridae
  genomes, marked in green, are more conserved than the others. This is
  due to the presence of overlapping reading frames covering on average
  30~\% of their genome. To see this figure in color, go online. \label{fig:mld}}
\end{figure}

This limited genome composition variability is further thinned down both by the imposed conservation of the dinucleotide  composition characteristic for the virus family and, in some viruses, by the presence of overlapping reading frames which  dramatically reduce the possibility to mutate the third nucleotide in a codon. Due to these two factors, it is found  that typical sequence identity  between WT sequences and their synonymous mutations ranges from about 66~\% to 85~\%, as shown in Fig.~\ref{fig:mld}(b).

The sequence space of synonymous mutations is thus so severely
restricted that there is no reason to expect that their progressive
accumulation has the same effect on compactness as the unrestricted
random shuffling of viral RNA sequences. As a matter of fact, the constrained synonymously-mutated
sequences could have, {\em a priori}, about the same compactness as
WT sequences or even improve it!
To support the earlier observations % conclusions
that WT RNAs are optimized for their spatial compactness, one must therefore
necessarily demonstrate that the accumulation of synonymous mutations, while leaving
the encoded protein phenotype and the chemical composition of the sequence unchanged, progressively destroys the
spatial compactness observed in WT sequences and quantified by their
respective MLDs.

To address this point, we start from WT viral RNA sequences and generate a mutation
flow in the sequence space using a Monte Carlo (MC) algorithm which proposes point mutations of the sequence and accepts or rejects them based on the constraints of synonymity and the conservation of the dinucleotide frequencies %strictly enforces the conservation of both dinucleotide frequencies and chemical composition
characteristic for a given virus family (see also Materials and Methods). The typical compactness of the resulting synonymously mutated WT genomes is again characterized by the asymptotic value of $\MLD$, averaged additionally over different mutated sequences and denoted by $\overline{\MLD}$.

The resulting MLDs are shown in Fig.~\ref{fig:mld}(a). It is indeed striking to
notice that despite the strongly reduced available sequence space, the $\overline{\MLD}$ of
synonymously mutated sequences falls on the same curve which describes the $\overline{\MLD}$ of
random sequences, given by the power law in Eq.~(\ref{eq:randMLD}). This fundamental observation can be condensed in the symbolic statement:
\begin{equation}
  \overline{\MLD}_{_{\rm WT}}(N) \underset{\rm (syn)}{\longrightarrow} \overline{\MLD}_{_{\rm random}}(N),
\label{foo1}
\end{equation}
\noindent where $N$ is the genome length and the arrow is a shorthand
for indicating the flow in the synonymous mutations subspace.

This result proves the conjecture that the WT genomes are indeed
characterized by a certain optimality of the MLD which, in turn,
reflects atypically-high degrees of RNA fold compactness. In fact, the
results of Fig.~\ref{fig:mld}(b) demonstrate that the WT MLD/compactness
can be obliterated even within a much restricted subset of mutations
that otherwise leave the viral phenotype and sequence
composition unchanged.
%Subsequently, one can conclude that the sequence phase space associated with synonymous mutations, despite being vanishingly small compared to the
%phase space of random mutations, is still large enough to allow for controlling
%the compactness and consequently also the packing characteristics
%of viral RNAs, as gauged by the MLD measure.

As an aside, we note that Tymoviridae exhibit an atypical behavior, with the
limiting value of $\overline{\MLD}$ under the synonymous mutation flow approaching
values which are still below the ones characteristic for random RNAs. The reason for this lies in the fact that  
Tymoviridae have a different nucleotide composition with respect to
other viral families; accounting for this different composition one
obtains a different prefactor for the scaling law in Eq.~(\ref{eq:randMLD}),
corresponding to more compact values of MLD, as shown in Fig.~\ref{fig:1}(c); see also SI Fig. S3 for more
details.

\subsection*{Synonymous mutation flow and the stability of genome MLD}

The previous result leads us to examine the details of the implied synonymous mutation flow 
[Eq.~(\ref{foo1})] and the stability of the terminal, asymptotic state of
the mutated sequence. In particular, we wish to establish the minimal
number of point nucleotide mutations that are needed to bring the MLD of
a viral RNA from its WT value to the random reference value. It is especially
interesting to ascertain whether this change in compactness happens
progressively, indicating that a continuous accumulation of mutations is
responsible for disrupting the WT RNA spatial compactness, or whether
the change is due to sporadic, punctuated events, which would suggest the
presence of specific RNA ``hotspots'' where mutations can dramatically
affect fold compactness.

To illuminate this point we considered 9 synthetic synonymous mutation
flow trajectories for 4 different viral sequences extracted from 3
viruses picked at random from 3 different families: brome mosaic virus
(BMV), ononsis yellow mosaic virus (OnYMV), and equine rhinitis B virus
1 (ERBV1). The considered sequences were chosen in order to probe the whole range of genome lengths
spanning from $N\simeq 2800$ nt to $N\simeq 8800$ nt. The trajectories
  were generated using the same MC scheme used to generate the equilibrium data
  presented in Fig.~\ref{fig:mld} (see also Materials and Methods), but 
 with a much more frequent sampling of the mutated sequences (every $N/100$ 
 attempted synonymous mutations) so as to leave detectable correlations in the
 series of generated sequences -- in this way mimicking the viral mutation dynamics.

\begin{figure}[h]
\begin{center}
\includegraphics[width=0.5\columnwidth]{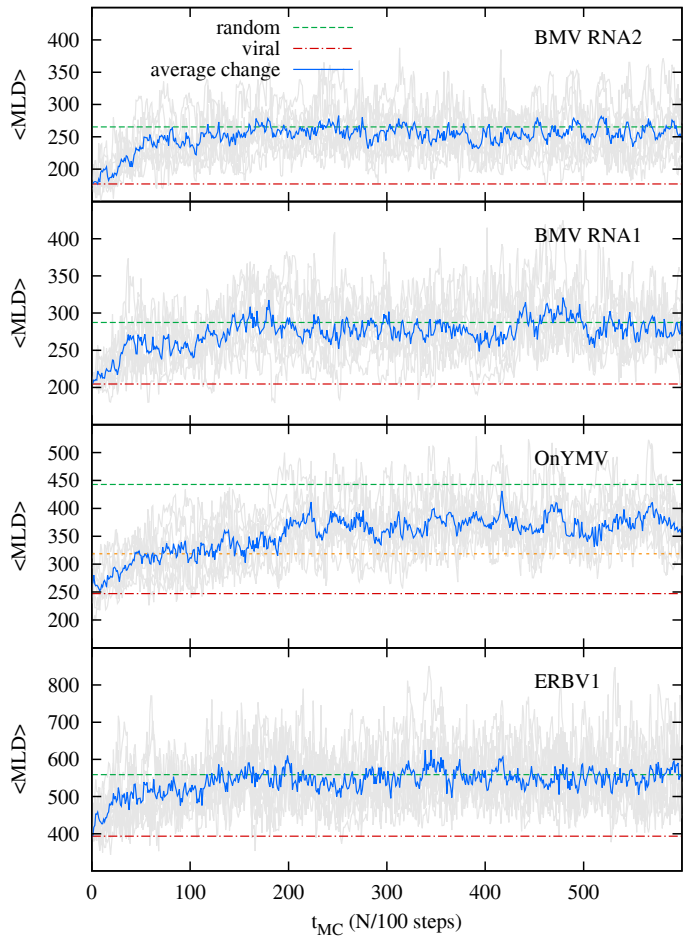}      %%color online
\end{center}
\caption{Mutation dynamics trajectories for 4 viral ssRNA sequences. From
 top to bottom: BMV RNA2 and RNA1 segments from the tripartite genome
  of BMV (Bromoviridae), OnYMV (Tymoviridae), and ERBV1 (Picornaviridae).
  Each panel shows 9 $\MLD$ trajectories and their average value (blue) for
  each sequence in units of MC steps, $N/100$. Red dot-dashed lines and
  green dashed lines show respectively the $\MLD$ values of WT RNAs and the $\overline{\MLD}$
    values of random RNAs [for viral-like composition,
  Eq.~(\ref{eq:randMLD})]. 
Notice that in the case of OnYMV, a Tymovirus, we must consider the appropriate asymptotic value of
$\overline{\MLD}$ for random RNAs with Tymoviridae-like composition (see Fig.~\ref{fig:1}). This value is
shown by the orange short-dashed line. To see this figure in color, go online.
\label{fig:evol_A}}
\end{figure}

The results are shown in Fig.~\ref{fig:evol_A}. From the mutation flow trajectories we discern that, at least for the sequences considered, the
change in compactness follows the continuous and gradual accumulation of
synonymous mutations, and does not take place in a punctuated
manner. Nonetheless, not many mutations are needed to make the MLD of
these sequences already indistinguishable from that of randomized RNAs.
In fact, mutating not more than $\sim 5$~\% of the full genome suffices
to erase the characteristic WT RNA compactness imprint.

A further interesting point clarified by the mutation flow trajectories
shown in Fig.~\ref{fig:evol_A} is that the genome fold compactness is
not completely optimized even in the case of WT sequences. In fact, for
all the 4 sequences considered in Fig.~\ref{fig:evol_A} one occasionally
observes more compact folded states, particularly during the initial part
of the trajectories.

\begin{figure}[h]
\begin{center}
\includegraphics[width=0.5\columnwidth]{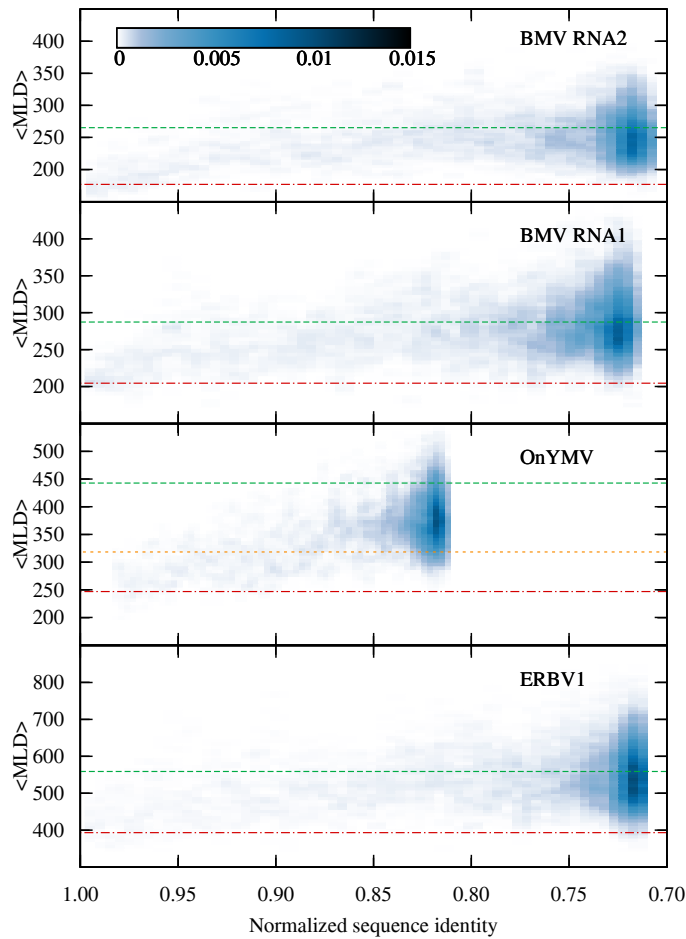}      %%color online
\end{center}
\caption{Color-coded heat maps for the probability density of finding
  mutated sequences with given $\MLD$ and sequence identity with the WT
sequence. The probability density for each virus is computed and normalized over the whole length of the $9$ mutation trajectories ($1500$ MC steps) shown in Fig.~\ref{fig:evol_A}.
Red dot-dashed lines and green dashed lines show respectively the $\MLD$ values of WT RNA and
the $\overline{\MLD}$ values of random RNAs [with
 viral-like composition, Eq.~(\ref{eq:randMLD})].  The orange short-dashed
 line in the OnYMV case shows the random $\overline{\MLD}$ value for Tymoviridae-like
composition. To see this figure in color, go online.
\label{fig:evol_B}}
\end{figure}

To better explore this interesting observation, we
computed the probability density of finding mutated sequences with given $\MLD$ 
as a function of the sequence identity to the WT sequence ratio, and plotted
it as a color-coded heatmap. These probability density plots are shown in
Fig.~\ref{fig:evol_B}, and we can observe that for some of the
genomes considered, such as BMV RNA1 and ERBV1, more compact structures are
reachable even when almost all the unconstrained nucleotides have already been
mutated. This point is most relevant in the present context. In fact, it 
demonstrates that the sequence-based synonymity constraint and the structure-based 
one for fold compactness, despite being in competition, can still be compatible.

This point is made more poignantly by considering the near-native pool of synonymous
sequences (e.g., those with sequence identity $\geq 95$~\%) for the
four cases presented in Fig.~\ref{fig:evol_B}. Across these instances it is found 
that from 12~\% to 21~\% of the near-native synonymous sequences 
have a predicted fold compactness that is equal or higher than the wild-type one.
This indicates that the well-optimised viral sequences still have a %substantial 
portion of phase space available for evolving while respecting both 
sequence- and structure-based stringent constraints. 
This appreciable residual mutation freedom may be clearly necessary to simultaneously
accommodate other concurrent selection constraints.

\subsection*{Taking into account codon usage bias and untranslated regions}

Finally, we examine the effect of two additional constraints which are
known to be relevant for some viruses, and may play a role in
maintaining viral RNA compactness. The first constraint is given by the
presence of functionally important secondary
RNA structures in the untranslated regions (UTRs) at the 3' and 5'
ends of several viral genomes~\cite{marz2014,tsukiyama1992,alvarez2005}. We take into account
this constraint by simply limiting the mutation flow to the coding
regions of the genomes.  Note that with this additional constraint our
theoretical estimate of the overall sequence identity
between WT sequences and sequences mutated asymptotically to saturation moves from
$66$-$73$~\% to $76$-$83$~\%. %\textcolor{red}{Mutated to saturation, means asymptotically? Not sure if wealso want to say that the more similar sequences are more heterogeneously-mutated (some parts intact, other changed).}

The second additional constraint is given by the fact that, since viruses adapt to
their hosts, not all the codons which translate into the same amino-acid
are statistically equivalent: some of them are more probable than others.
This codon usage bias is known to be an important
constraint for several viruses. In fact, changing the codon bias or the
codon-pair bias leads to attenuated viruses and has been proposed as a
possible vaccination strategy~\cite{Bull2012,Coleman2008}. To produce
mutated sequences with WT codon populations we shuffled the equivalent codons within
every viral gene (see Materials and Methods for details regarding the
implementation of codon-bias preserving synonymous mutations).%, as described in Materials and Methods. 

\begin{figure}[h]
\begin{center}
\includegraphics[width=0.5\columnwidth]{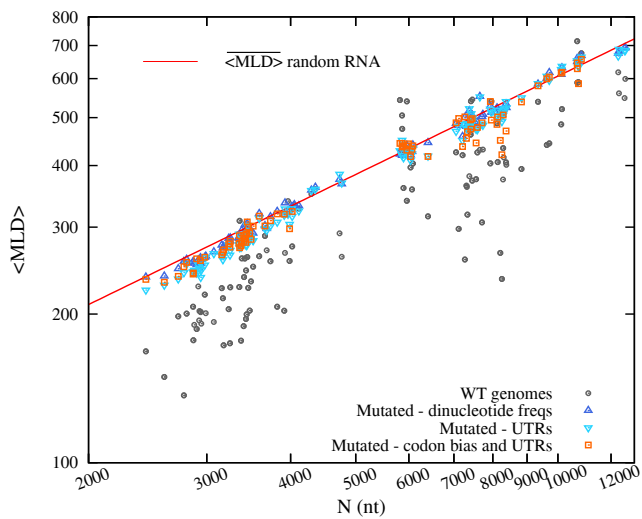}   %% color online
\end{center}
\caption{The $\overline{\MLD}$ values for  the 
  synonymous constraint only (upward triangles) and for the additional constraints of preserving UTRs sequences
  (downward triangles) and UTRs sequences as well as codon biases (squares).
  The $\overline{\MLD}$ values for these last two cases are evaluated over a set of
  $150$ mutated sequences for each virus.
  Data are presented in the same manner as in Fig.~\ref{fig:mld} 
  (see also SI Fig. S3 for UTRs preserving synonymous point mutations of
Tymoviridae). To see this figure in color, go online. \label{fig:cbs}}
\end{figure}

%\textcolor{red}{As it is formulated it means that the shuffling of equivalent codon is done only within single genes. This is not the same as substituting equivalent codons (in the coding regions) at a genome-wide scale. Please check if it is OK as written!}. 
%%This is indeed the case. Codon biases appear in general to depend on the
%%expression levels of single genes. Those more optimized being translated
%%more often. I therefore played it safe and implemented it at a single gene
%%level.
The results obtained with both of these constrains are compared in
Fig.~\ref{fig:cbs} against
those previously obtained using synonymous point mutations. It is important to notice that even with these
additional constraints, which further thin out the phase
space available to mutations, our results remain valid, confirming the
presence of an evolutionary pressure to produce compact RNA folds.

\section*{DISCUSSION AND CONCLUSIONS}\label{sec:disc}

While the fundamental mechanisms by which point mutations affect 
the fitness of the organisms in their respective environments
({\em via} the transcription of the mutated nucleotide sequence into the modified protein products) 
are well understood~\cite{Wylie:2011:PNAS,Chen:2009:Genetics,Duffy2008}, it is less known what are 
their effects on the purely physico-chemical properties of their genomes. 
In order to investigate possible parallel selection mechanisms and eventual embedded levels
of coding that control the compactness of viral ssRNA folds, we analyzed a synthetic
model for accumulating synonymous mutations in viral RNAs and assessed their impact on the spatial compactness of the genome as quantified by the MLD measure, introduced by Yoffe and coworkers~\cite{Yoffe2008}. 
We have analyzed the effects of synonymous mutations under
different constraints on ssRNA genomes for a large number of different viral families with icosahedral capsids, 
and compared the changes in their compactness with randomly shuffled RNA sequences 
with the same nucleotide composition, which are in general significantly less compact than those encapsidated by viruses.

Using extensive computational analysis we have shown that progressive accumulation of synonymous point mutations, although neutral from the
functional point of view as they conserve the expressed protein
complement, completely erases the typical compactness of viral WT RNA
folds. In fact, under the synonymous
mutation flow the MLDs of WT RNAs approach their corresponding random RNA values in a
continuous manner even after a relatively small number of
mutations. Although, in principle, the emergence of viral RNA fold compactness may still be related
to some other evolutionary pressure, our results rule out the principal
ones, including codon bias and the preservation of
functional UTRs, and thus strongly support the independent evolution of viral
RNA fold compactness. Arguably, such a dramatic reduction in RNA fold compactness,
which in this respect eventually makes it undistinguishable from a
random RNA sequence, has a relevant impact on the virion assembly and
therefore on the ability of viruses to replicate and propagate their
infection. %We furthermore showed that the typical WT RNA
These results are strengthened by the observation that the typical WT RNA
  compactness is not related to codon usage bias nor is it dictated by the
  particular sequence/structure of its non-coding regions, since synonymous mutations which
maintain both unchanged were found to nonetheless destroy the typical WT RNA compactness.

The connection between the viral %genomic 
RNA sequence and its
physical properties, such as its compactness, 
may in future allow to control the physical properties of viral RNAs and specifically their
aptitude for efficient packing. This we believe may lead to improve and broaden the scope 
of existing strategies which harness viral mutation rates to achieve virus
attenuation.

%\begin{acknowledgments}
\section*{SUPPORTING MATERIAL}
Seven figures and two tables of supporting data.

LT, ALB, and RP acknowledge support from ARRS Grants No. P1-0055 and J1-4297. CM acknowledges support from the Italian Ministry of Education, grant PRIN No. 2010HXAW77.
%\end{acknowledgments}

%\bibliographystyle{biophysj}
%\bibliographystyle{BJ}
%\bibliography{BJ_references}\end{document}

\newpage
\setcounter{figure}{0}
\renewcommand{\figurename}{SI Fig.}%
\renewcommand{\tablename}{SI Table}%
\input{Tubiana_et_al_Synonymous_Mutations_Erase_Viral_RNA_Compactness_SI.tex}

\end{document}

%% file: Tubiana_et_al_Synonymous_Mutations_Erase_Viral_RNA_Compactness_SI.tex
\vspace{0.5cm}
\centerline{\huge{Supporting Information}}
\vspace{0.5cm}

\section{Fit of the shuffled RNA MLD}

To obtain the power law for the MLD of random RNAs, we shuffled $12$ RNA
sequences of different lengths (1000 nt, 1500 nt, \ldots, 6000 nt), all
having a viral-like nucleotide composition: 0.26 A,  0.28 U, 0.24  G, 0.22 C
(obtained excluding Tymoviridae, which have a significantly different
composition). For every sequence length, we produced 500 independent
sequences over which we computed the expected (thermally averaged) $\MLD$.
The power law of Eq. (1) in the main text is then obtained by fitting the
dependence of $\overline{\MLD}$, further averaged over the 500 different mutations, on the sequence length. 

As already mentioned in the main text, Tymoviridae differ notably from the
other families in their nucleotide composition, and they were not considered
when producing the averaged viral-like composition. Evaluating the average
composition for the set of Tymoviridae viruses considered in the main text,
we obtain  0.20  A, 0.24 U,  0.18  G, 0.38 C.
%we obtain: 0.219 A, 0.254 U, 0.163 G, 0.364 C. 

\begin{figure}[b]
\centering
\includegraphics[width=0.8\textwidth]{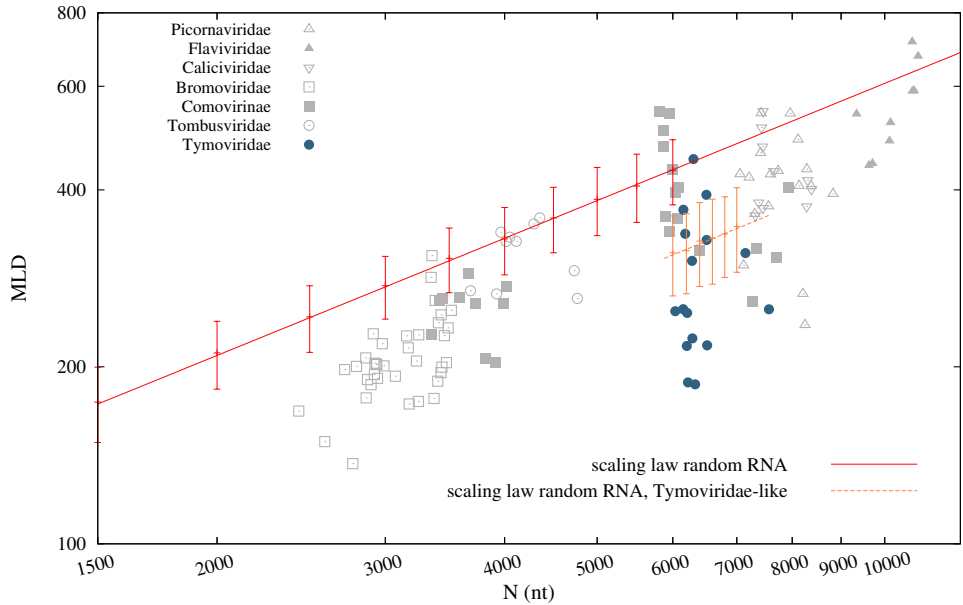}
\caption{
$\MLD$ values of WT RNA genomes are shown in gray for all families apart from Tymoviridae, which are highlighted in blue.
$\overline{\MLD}$ values of random sequences are shown with red and orange errorbars for viral-like
and Tymoviridae-like nucleotide composition, respectively. The respective fitting lines are displayed with the same colors.
The $p$-value of the fit parameters for the viral-like composition is below $10^{-10}$, and the adjusted $R^2$ is $0.999948$. 
For Tymoviridae-like composition the $p$-value of exponent is $\simeq 10^{-4}$ and the adjusted $R^2$ is $0.999968$.
\label{fig:SI_fig1}}
\end{figure}

Using this alternative composition and adopting the same procedure used for the other families 
we obtain a scaling law describing the $\overline{\MLD}$ dependence of Tymoviridae-like random RNA sequences:  
\begin{equation}
%MLD_{\mathrm{Ty}}(N) = (0.87\pm 0.34)\times N^{(0.669\pm 0.039)}.
  \overline{\MLD}_{\mathrm{Ty}}(N) = (0.92\pm0.44)\times N^{(0.669\pm 0.054)}.
\label{eq:SI_1}
\end{equation}
Note that the exponent, $0.669 \pm 0.054$, is compatible with the one obtained for the other viral families, $0.662\pm 0.004$. Both fits are shown in Fig.~S\ref{fig:SI_fig1}.

\begin{figure}[t]
\centering
\includegraphics[width=0.8\textwidth]{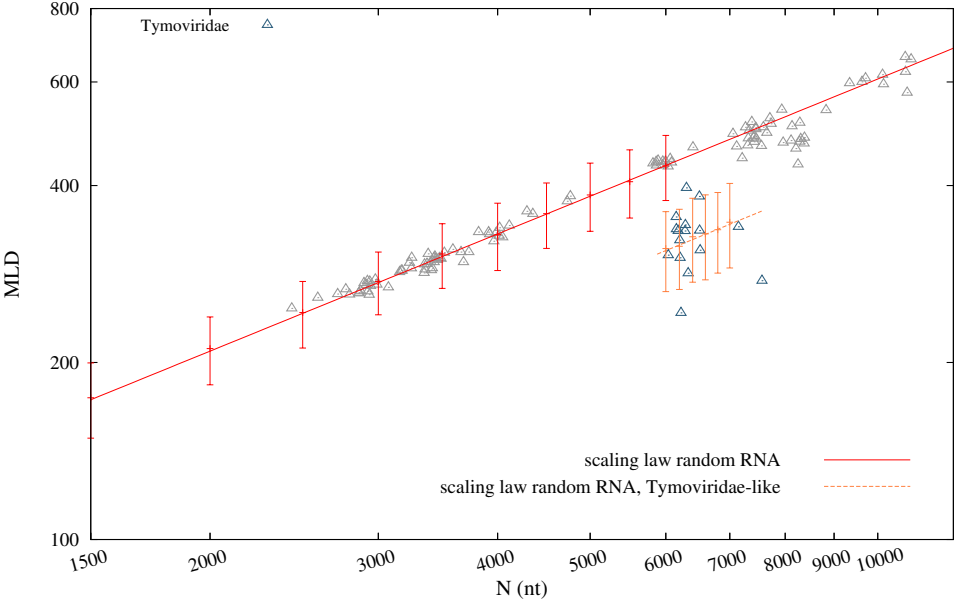}
\caption{
$\MLD$ values of randomly shuffled WT RNA genomes, shown in gray for all families apart from Tymoviridae, which are highlighted in blue.
$\overline{\MLD}$ values of random sequences are shown with red and orange errorbars for viral-like
and Tymoviridae-like nucleotide composition, respectively. 
\label{fig:SI_fig2}}
\end{figure}

We further check the validity of the scaling laws for our viral families by
randomly shuffling the WT RNA sequences themselves, without any further
constraints. The results, shown in Fig.~S\ref{fig:SI_fig2}, show once again that the two
scaling laws are a good reference for random RNAs with the viral-like composition
considered in our sample. For Tymoviridae, we notice that a couple of
viruses remain more compact than predicted by Eq.~(S\ref{eq:SI_1}). This is due
to them having a composition which is substantially different from the
Tymoviridae average composition.

%Accounting for the different scaling law for Tymoviridae, we can recover our
%main results obtained in the main text for the other viral families also for
%the case of Tymoviridae. In fact, as shown in Fig.~\ref{fig:SI_fig3},  
%we can see that the limiting values of MLD for Tymoviridae under the influence of the synonymous mutation flow are even higher than what would be expected from the scaling law of Eq.~\ref{eq:SI_1}.

%This is arguably due to the fact that even if dinucleotide frequencies are conserved,
%nucleotide frequencies can still vary slightly, see Fig.~\ref{fig:SI_fig4}. Indeed, if instead of
%conserving dinucleotide frequencies one enforces chemical composition by
%simply shuffling the sequences, again under the constraint of conserved
%amino acid sequence, the MLDs of the mutated Tymoviridae sequences are
%more compatible with the prediction of Eq.~(\ref{eq:SI_1}), as shown in
%Fig.~\ref{fig:SI_fig3}. Our general claim, though, remains unaffected.

\section{Mutations preserving UTRs}
\begin{figure}[!h]
\centering
\includegraphics[width=0.7\textwidth]{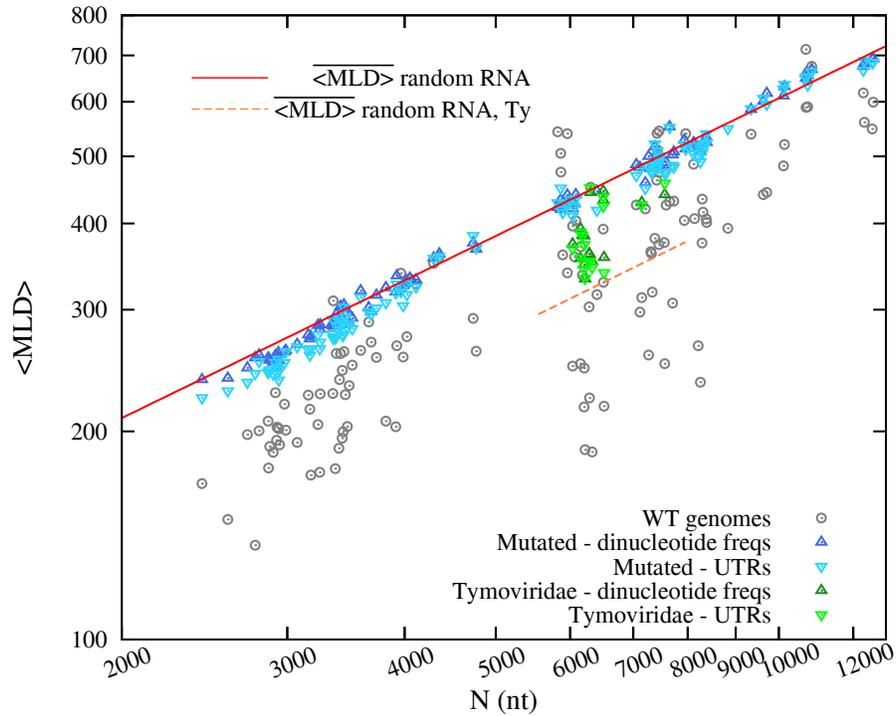}
\caption{Comparison between the $\overline{\MLD}$ values for  the 
  synonymous constraint only (upward triangles) and for the additional constraints of preserving UTRs sequences
  (downward triangles), including Tymoviridae. In the latter case
  $\overline{\MLD}$ have been evaluated over a set of $150$ mutated
  sequences per virus.  \label{fig:SI_UTRs}}
\end{figure}

As detailed in the main text, we further tested the robustness of our
results by adding additional optional constraint as the preservation of
Untranslated regions (UTRs) near the ends of the genome and the preservation
of the codon biases within each gene. The $\overline{\MLD}$ values obtained 
with these additional constraint are compared with those obtained under the
constraint of synonymous mutations only  in Fig. 5 in the main text. Here, in
Fig.~S\ref{fig:SI_UTRs} we extend the comparison for the additional
constraint of preserving UTRs to include Tymoviridae. $\overline{\MLD}$
values under the additional constraint of fixed codon bias were not
calculated for this family since all the tymoviridae genomes in our set present
overlapping genes.

\section{Mutations at fixed nucleotide composition}
\begin{figure}[!h]
\centering
\includegraphics[width=0.8\textwidth]{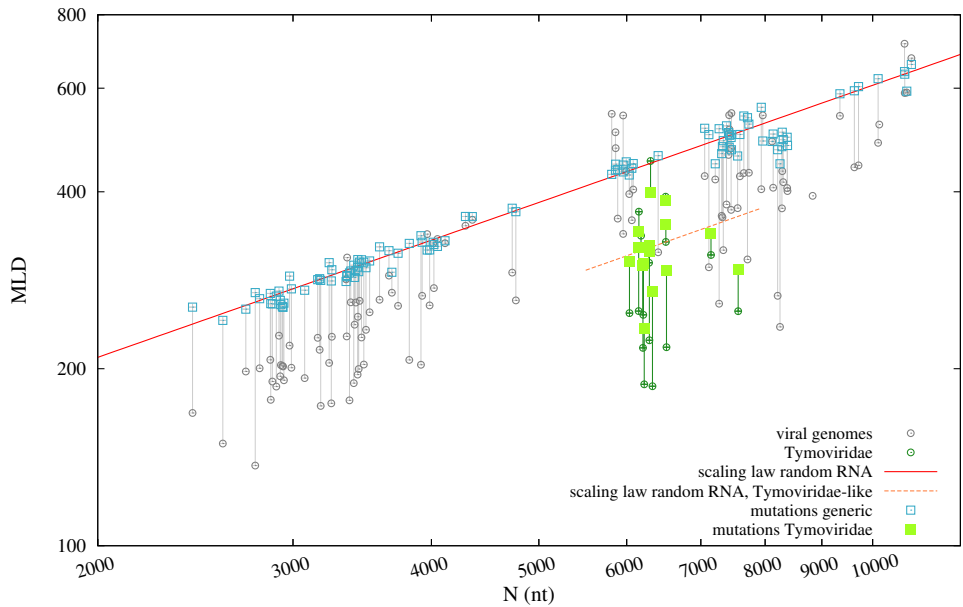}
\caption{Mutations performed at fixed nucleotide composition. Note that the
  $\overline{\MLD}$ of the mutated viral sequences approaches the random RNA values for viral-like
  and Tymoviridae-like nucleotide composition in both respective cases. We note
  that for Tymoviridae there are some viruses which remain more compact than
  the corresponding random RNAs. We argue that this is due to the fact that
  Tymoviridae show notable fluctuations in their nucleotide composition.
\label{fig:SI_fig3}}
\end{figure}
To test the robustness of the results reported in the main text, we
implemented another mutation flow which conserves the nucleotide composition instead of the
dinucleotide frequencies. This is achieved by using a Fisher-Yates algorithm where
proposed shuffles are accepted or rejected on the basis of whether or not
the resulting genome still encodes for the same proteins. The results of this
different simulation setup are shown in Fig.~S\ref{fig:SI_fig3}.

Note that the values of $\overline{\MLD}$ obtained in this way show a clear correlation
with those obtained by unrestricted random shuffling of the WT RNA
sequences, shown in Fig.~S\ref{fig:SI_fig2}.

\section{Details of dinucleotide and nucleotide compositions}

\begin{figure}[!h]
\centering
\includegraphics[width=\textwidth]{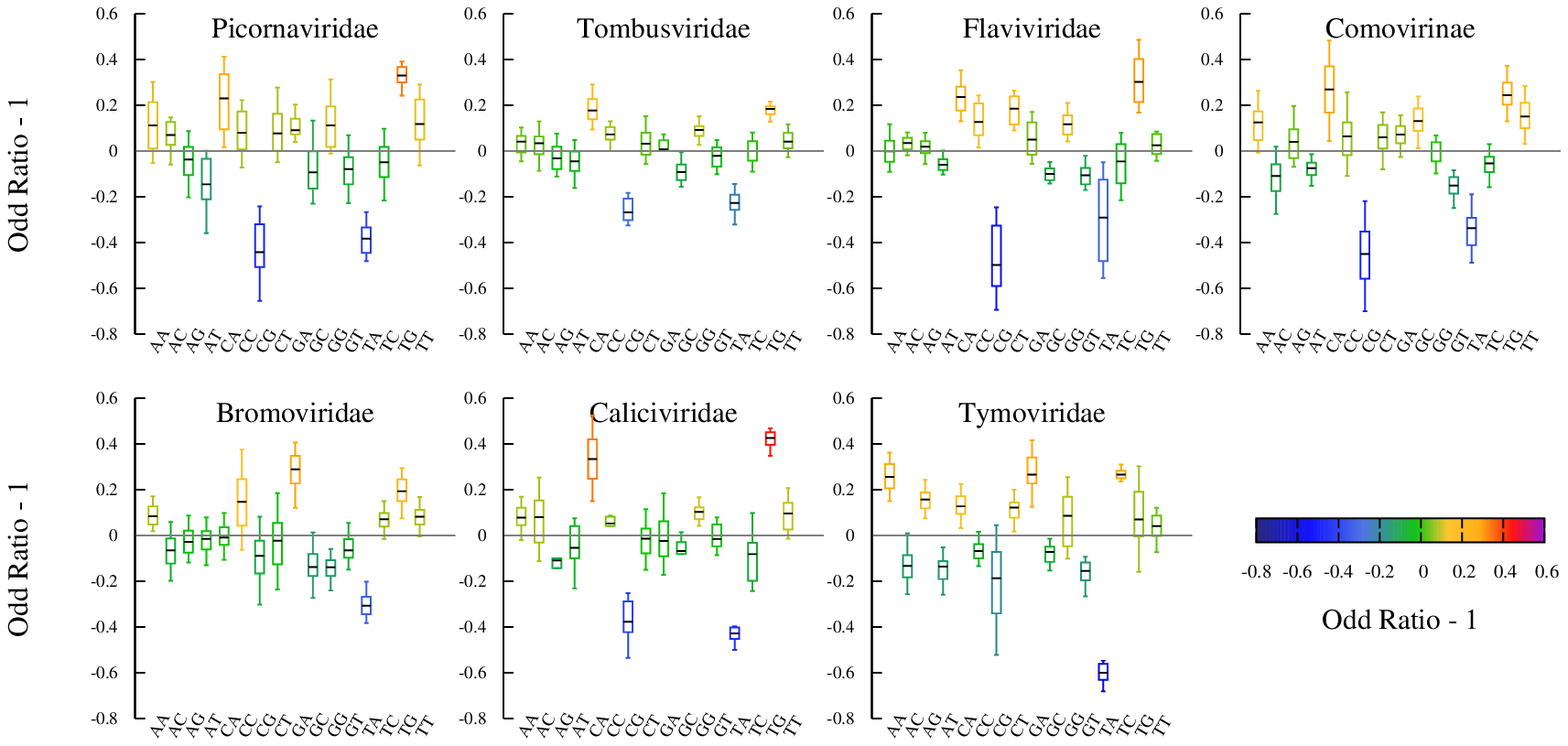}
\caption{ Dinucleotide odd-ratios, rescaled to zero, for the viral families considered in
  our study. Boxes represent  quartiles, and whiskers correspond to $1.5$ of
  the interquartile distance. These values have been used to constrain the
  mutation flow and produce sequences with viral-like dinucleotide
  frequencies, see Materials and Methods section in the main text.
\label{fig:SI_fig4}}
\end{figure}

\begin{figure}[!h]
\centering
\includegraphics[width=\textwidth]{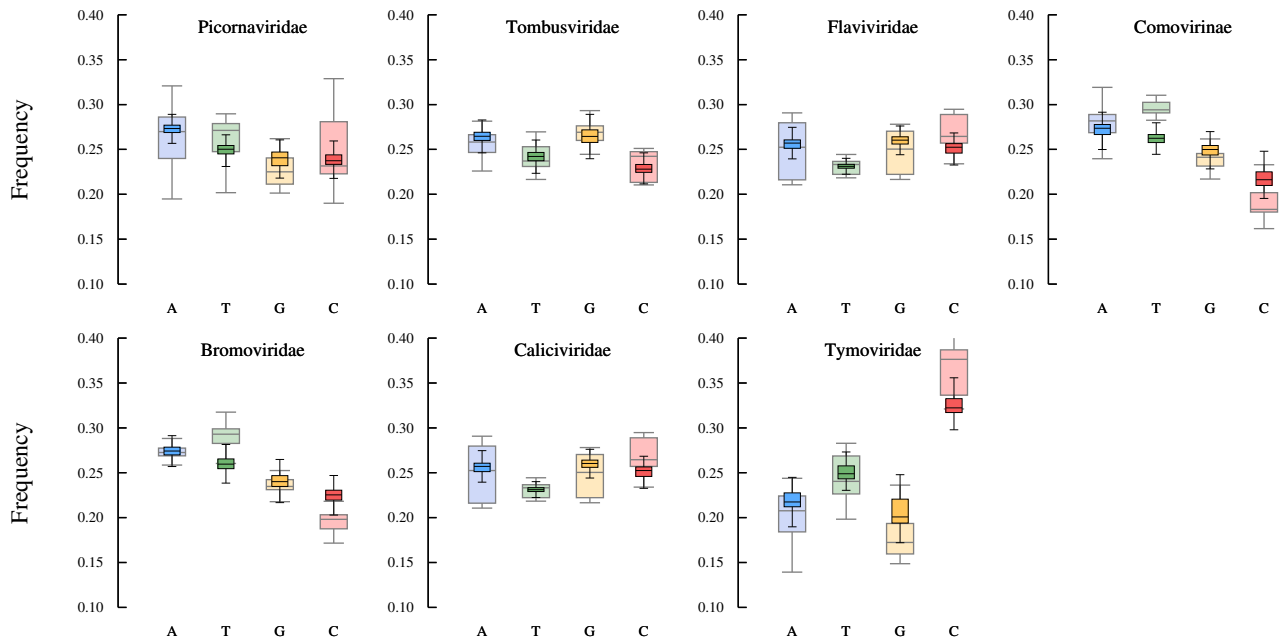}
\caption{Nucleotide frequencies for WT sequences (large boxes) and mutated sequences with constrained dinucleotide composition (small boxes), shown for each virus family considered in our study. Note that in most families the imposition of conserved dinucleotide frequencies results in conserved nucleotide frequencies as well, although for Tymoviridae, Comovirinae, and Bromoviridae the frequencies are not so well preserved, showing the effects of transversion changes.\label{fig:SI_fig5}}
\end{figure}

\newpage\null\thispagestyle{empty}\newpage

\section{Dataset of viral genomes}
  \centering
  \LTcapwidth=\textwidth
  \begin{longtable}{llllcccccc}
  \bf{Taxon}&\bf{Family}&\bf{Nuccore code}&\bf{ PDB code} &\bf{ length}
    &\bf{ $\MLD_{WT}$}
    &\bf{$\overline{\MLD_{mut}}$}&\bf{$\overline{\MLD_{UTRs}}$}&\bf{$\overline{\MLD_{CB}}$}\\
  \cmidrule{1-10}\\
  Bromoviridae      &	Anulavirus     &	\verb;PeZSV_RNA1      ;       &	--&	     3383&  $259\pm19$&  $295\pm48$&  $274\pm45$&   $280\pm43$\\
Bromoviridae      &	Anulavirus     &	\verb;PeZSV_RNA2      ;       &	--&	     2435&  $168\pm8 $&  $238\pm39$&  $224\pm37$&   $235\pm38$\\
Bromoviridae      &	Bromovirus     &	\verb;BMV_RNA1        ;       &	1js9&	     3234&  $204\pm15$&  $285\pm46$&  $276\pm43$&   $278\pm46$\\
Bromoviridae      &	Bromovirus     &	\verb;BMV_RNA2        ;       &	1js9&	     2865&  $177\pm12$&  $255\pm42$&  $244\pm38$&   $242\pm40$\\
Bromoviridae      &	Bromovirus     &	\verb;BrBMV_RNA1      ;       &	--&	     3158&  $225\pm21$&  $276\pm44$&  $263\pm43$&   $264\pm42$\\
Bromoviridae      &	Bromovirus     &	\verb;BrBMV_RNA2      ;       &	--&	     2799&  $200\pm17$&  $258\pm43$&  $252\pm40$&   $255\pm43$\\
Bromoviridae      &	Bromovirus     &	\verb;CaYBV_RNA1      ;       &	--&	     3178&  $172\pm14$&  $274\pm45$&  $263\pm40$&   $270\pm45$\\
Bromoviridae      &	Bromovirus     &	\verb;CaYBV_RNA2      ;       &	--&	     2720&  $197\pm20$&  $247\pm41$&  $236\pm40$&   $239\pm39$\\
Bromoviridae      &	Bromovirus     &	\verb;CCMV_RNA1       ;       &	1cwp&	     3171&  $215\pm27$&  $272\pm44$&  $258\pm40$&   $267\pm44$\\
Bromoviridae      &	Bromovirus     &	\verb;CCMV_RNA2       ;       &	1cwp&	     2774&  $136\pm11$&  $256\pm43$&  $243\pm38$&   $249\pm40$\\
Bromoviridae      &	Bromovirus     &	\verb;MeYFV_RNA1      ;       &	--&	     3249&  $174\pm19$&  $281\pm46$&  $263\pm44$&   $274\pm47$\\
Bromoviridae      &	Bromovirus     &	\verb;MeYFV_RNA2      ;       &	--&	     2862&  $207\pm11$&  $255\pm40$&  $243\pm38$&   $241\pm37$\\
Bromoviridae      &	Bromovirus     &	\verb;SpBLV_RNA1      ;       &	--&	     3252&  $226\pm25$&  $285\pm45$&  $269\pm43$&   $281\pm47$\\
Bromoviridae      &	Bromovirus     &	\verb;SpBLV_RNA2      ;       &	--&	     2898&  $186\pm16$&  $252\pm41$&  $242\pm40$&   $259\pm39$\\
Bromoviridae      &	Cucumovirus    &	\verb;GaMMV_RNA1      ;       &	--&	     3350&  $283\pm28$&  $286\pm47$&  $277\pm43$&   $278\pm44$\\
Bromoviridae      &	Cucumovirus    &	\verb;GaMMV_RNA2      ;       &	--&	     2935&  $202\pm8 $&  $260\pm43$&  $254\pm39$&   $257\pm42$\\
Bromoviridae      &	Cucumovirus    &	\verb;PeSV_RNA1       ;       &	--&	     3357&  $309\pm16$&  $287\pm46$&  $273\pm44$&   $273\pm42$\\
Bromoviridae      &	Cucumovirus    &	\verb;TAV_RNA1        ;       &	1laj&	     3410&  $237\pm18$&  $287\pm46$&  $285\pm44$&   $281\pm49$\\
Bromoviridae      &	Cucumovirus    &	\verb;TAV_RNA2        ;       &	1laj&	     3074&  $192\pm18$&  $267\pm43$&  $265\pm42$&        -- \\
Bromoviridae      &	Ilarvirus      &	\verb;ApMV_RNA1       ;       &	--&	     3476&  $203\pm36$&  $297\pm49$&  $283\pm45$&   $292\pm49$\\
Bromoviridae      &	Ilarvirus      &	\verb;ApMV_RNA2       ;       &	--&	     2979&  $218\pm20$&  $261\pm43$&  $251\pm43$&   $261\pm45$\\
Bromoviridae      &	Ilarvirus      &	\verb;CiLRV_RNA1      ;       &	--&	     3404&  $189\pm27$&  $289\pm46$&  $281\pm48$&   $289\pm 4$\\
Bromoviridae      &	Ilarvirus      &	\verb;CiLRV_RNA2      ;       &	--&	     2990&  $200\pm21$&  $262\pm43$&  $261\pm42$&        -- \\
Bromoviridae      &	Ilarvirus      &	\verb;CiVV_RNA1       ;       &	--&	     3433&  $245\pm17$&  $291\pm48$&  $287\pm45$&   $290\pm42$\\
Bromoviridae      &	Ilarvirus      &	\verb;CiVV_RNA2       ;       &	--&	     2914&  $227\pm29$&  $257\pm41$&  $252\pm40$&        -- \\
Bromoviridae      &	Ilarvirus      &	\verb;ElMV_RNA1       ;       &	--&	     3431&  $195\pm11$&  $285\pm46$&  $276\pm41$&   $279\pm44$\\
Bromoviridae      &	Ilarvirus      &	\verb;ElMV_RNA2       ;       &	--&	     2874&  $190\pm25$&  $254\pm41$&  $246\pm43$&        -- \\
Bromoviridae      &	Ilarvirus      &	\verb;ParMV_RNA1      ;       &	--&	     3518&  $249\pm20$&  $292\pm48$&  $282\pm44$&   $301\pm50$\\
Bromoviridae      &	Ilarvirus      &	\verb;ParMV_RNA2      ;       & --&	     2922&  $194\pm21$&  $247\pm40$&  $248\pm39$&        -- \\
Bromoviridae      &	Ilarvirus      &	\verb;PrDV_RNA1       ;       &	--&	     3374&  $176\pm24$&  $285\pm46$&  $273\pm42$&   $275\pm47$\\
Bromoviridae      &	Ilarvirus      &	\verb;PrDV_RNA2       ;       &	--&	     2593&  $149\pm17$&  $239\pm39$&  $229\pm37$&   $232\pm39$\\
Bromoviridae      &	Ilarvirus      &	\verb;SpLV_RNA1       ;       &	--&	     3439&  $199\pm19$&  $291\pm48$&  $275\pm44$&   $291\pm46$\\
Bromoviridae      &	Ilarvirus      &	\verb;SpLV_RNA2       ;       &	--&	     2939&  $201\pm22$&  $253\pm40$&  $237\pm37$&        -- \\
Bromoviridae      &	Ilarvirus      &	\verb;ToSV_RNA1       ;       &	--&	     3491&  $232\pm28$&  $286\pm46$&  $286\pm48$&   $283\pm46$\\
Bromoviridae      &	Ilarvirus      &	\verb;ToSV_RNA2       ;       &	--&	     2926&  $202\pm15$&  $253\pm42$&  $243\pm40$&        -- \\
Bromoviridae      &	Ilarvirus      &	\verb;TuAMV_RNA1      ;       &	--&	     3459&  $226\pm17$&  $301\pm48$&  $292\pm47$&   $300\pm48$\\
Bromoviridae      &	Ilarvirus      &	\verb;TuAMV_RNA2      ;       &	--&	     2944&  $191\pm9 $&  $258\pm41$&  $246\pm40$&        -- \\
Caliciviridae     &	Nebovirus      &	\verb;caliciNB        ;       & --&	     7453&  $473\pm48$&  $502\pm79$&  $501\pm77$&   $498\pm83$\\%
Caliciviridae     &	Nebovirus      &	\verb;newbury         ;       &	--&	     7454&  $372\pm18$&  $495\pm78$&  $496\pm82$&   $498\pm83$\\
Caliciviridae     &	Norovirus      &	\verb;murineNoro1     ;       &	--&	     7382&  $380\pm36$&  $517\pm81$&  $521\pm82$&   $491\pm83$\\
Caliciviridae     &	Norovirus      &	\verb;norwalk         ;       &	1ihm&	     7654&  $430\pm35$&  $552\pm84$&  $551\pm77$&        -- \\
Caliciviridae     &	Sapovirus      &	\verb;porcineSapo     ;       &	--&	     7320&  $361\pm36$&  $480\pm77$&  $486\pm73$&        -- \\
Caliciviridae     &	Sapovirus      &	\verb;sapoMc10        ;       &	--&	     7458&  $544\pm39$&  $486\pm78$&  $491\pm73$&        -- \\
Caliciviridae     &	Sapovirus      &	\verb;saporo          ;       &	--&	     7429&  $510\pm33$&  $508\pm79$&  $509\pm79$&        -- \\
Caliciviridae     &	Vesivirus      &	\verb;rabbitVV        ;       &	--&	     8380&  $401\pm20$&  $524\pm81$&  $523\pm82$&        -- \\
Caliciviridae     &	Vesivirus      &	\verb;stellerVV       ;       & --&	     8305&  $415\pm16$&  $508\pm79$&  $521\pm77$&        -- \\%
Caliciviridae     &	Vesivirus      &	\verb;VESV            ;       & --&	     8284&  $374\pm41$&  $516\pm76$&  $516\pm78$&        -- \\%
Comovirinae       &	Comovirus      &	\verb;BPMV_RNA1       ;       &	1bmv&	     5995&  $433\pm40$&  $430\pm67$&  $434\pm72$&   $443\pm66$\\
Comovirinae       &	Comovirus      &	\verb;BPMV_RNA2       ;       &	1bmv&	     3662&  $288\pm26$&  $302\pm49$&  $298\pm48$&   $302\pm50$\\
Comovirinae       &	Comovirus      &	\verb;CowSMV_RNA1     ;       &	--&	     5957&  $339\pm28$&  $427\pm69$&  $425\pm63$&   $430\pm62$\\
Comovirinae       &	Comovirus      &	\verb;CowSMV_RNA2     ;       &	--&	     3732&  $255\pm30$&  $315\pm51$&  $302\pm49$&   $309\pm54$\\
Comovirinae       &	Comovirus      &	\verb;CPMV_RNA1       ;       &	1ny7&	     5889&  $360\pm24$&  $423\pm66$&  $415\pm66$&   $439\pm72$\\
Comovirinae       &	Comovirus      &	\verb;RadMV_RNA1      ;       &	--&	     6064&  $357\pm21$&  $427\pm67$&  $422\pm63$&   $431\pm73$\\
Comovirinae       &	Comovirus      &	\verb;RadMV_RNA2      ;       &	--&	     4020&  $274\pm20$&  $329\pm53$&  $315\pm52$&   $323\pm50$\\
Comovirinae       &	Comovirus      &	\verb;RCMV_RNA1       ;       &	rcmv&	     6033&  $396\pm28$&  $420\pm65$&  $410\pm59$&   $417\pm61$\\
Comovirinae       &	Comovirus      &	\verb;SquashMV_RNA1   ;       &	--&	     5865&  $474\pm27$&  $419\pm69$&  $419\pm70$&   $436\pm71$\\
Comovirinae       &	Comovirus      &	\verb;SquashMV_RNA2   ;       &	--&	     3354&  $226\pm17$&  $285\pm48$&  $291\pm49$&   $288\pm45$\\
Comovirinae       &	Comovirus      &	\verb;TurRV_RNA1      ;       &	--&	     6082&  $403\pm32$&  $440\pm70$&  $434\pm70$&   $439\pm63$\\
Comovirinae       &	Comovirus      &	\verb;TurRV_RNA2      ;       &	--&	     3985&  $256\pm18$&  $325\pm52$&  $304\pm49$&   $298\pm46$\\
Comovirinae       &	Fabavirus      &	\verb;BBWV_RNA1       ;       &	--&	     5817&  $542\pm37$&  $422\pm68$&  $428\pm64$&   $444\pm74$\\
Comovirinae       &	Fabavirus      &	\verb;BBWV_RNA2       ;       &	--&	     3446&  $260\pm24$&  $305\pm49$&  $303\pm46$&   $307\pm49$\\
Comovirinae       &	Fabavirus      &	\verb;mikaniaMMV_RNA1 ;       &	--&	     5862&  $505\pm46$&  $433\pm69$&  $450\pm67$&   $443\pm73$\\
Comovirinae       &	Fabavirus      &	\verb;mikaniaMMV_RNA2 ;       &	--&	     3418&  $259\pm30$&  $303\pm49$&  $289\pm47$&   $285\pm51$\\
Comovirinae       &	Fabavirus      &	\verb;patchMMV_RNA1   ;       &	--&	     5956&  $539\pm24$&  $440\pm70$&  $428\pm68$&   $438\pm62$\\
Comovirinae       &	Fabavirus      &	\verb;patchMMV_RNA2   ;       &	--&	     3591&  $262\pm32$&  $320\pm51$&  $313\pm51$&   $316\pm55$\\
Comovirinae       &	Nepovirus      &	\verb;arabisMV_RNA1   ;       &	--&	     7334&  $318\pm30$&  $485\pm74$&  $475\pm78$&   $468\pm73$\\
Comovirinae       &	Nepovirus      &	\verb;arabisMV_RNA2   ;       &	--&	     3820&  $207\pm20$&  $323\pm53$&  $307\pm47$&   $319\pm45$\\
Comovirinae       &	Nepovirus      &	\verb;blackCRV_RNA1   ;       &	--&	     7711&  $306\pm31$&  $502\pm75$&  $486\pm76$&   $488\pm78$\\
Comovirinae       &	Nepovirus      &	\verb;blackCRV_RNA2   ;       &	--&	     6405&  $315\pm22$&  $445\pm67$&  $418\pm65$&   $417\pm71$\\
Comovirinae       &	Nepovirus      &	\verb;raspRV_RNA1     ;       &	--&	     7935&  $404\pm39$&  $528\pm83$&  $520\pm81$&   $539\pm81$\\
Comovirinae       &	Nepovirus      &	\verb;raspRV_RNA2     ;       &	--&	     3914&  $203\pm13$&  $318\pm50$&  $317\pm53$&   $319\pm48$\\
Comovirinae       &	Nepovirus      &	\verb;TRSV_RNA2       ;       &	1a6c&	     7271&  $257\pm20$&  $500\pm80$&  $484\pm75$&   $502\pm84$\\
Flaviviridae      &	Flavivirus     &	\verb;alkhurma        ;       &	--&	    10685&  $714\pm36$&  $659\pm10$&  $651\pm10$&   $646\pm96$\\
Flaviviridae      &	Flavivirus     &	\verb;apoi            ;       &	--&	    10116&  $484\pm38$&  $612\pm96$&  $626\pm97$&   $615\pm88$\\
Flaviviridae      &	Flavivirus     &	\verb;dengue          ;       &	--&	    10735&  $589\pm45$&  $654\pm99$&  $634\pm98$&   $586\pm96$\\
Flaviviridae      &	Flavivirus     &	\verb;montana         ;       &	--&	    10690&  $588\pm34$&  $649\pm99$&  $652\pm10$&   $628\pm92$\\
Flaviviridae      &	Flavivirus     &	\verb;powassan        ;       &	--&	    10839&  $674\pm52$&  $668\pm10$&  $663\pm97$&   $656\pm95$\\
Flaviviridae      &	Flavivirus     &	\verb;rioBravo        ;       &	--&	    10140&  $520\pm79$&  $631\pm10$&  $635\pm98$&   $618\pm91$\\
Flaviviridae      &	Hepacivirus    &	\verb;HepC2           ;       &	--&	     9711&  $443\pm36$&  $617\pm10$&  $595\pm86$&   $604\pm96$\\
Flaviviridae      &	Hepacivirus    &	\verb;HepC5           ;       &	--&	     9343&  $538\pm88$&  $585\pm97$&  $586\pm91$&   $580\pm86$\\
Flaviviridae      &	Hepacivirus    &	\verb;HepC6           ;       &	--&	     9628&  $440\pm25$&  $601\pm92$&  $607\pm93$&   $599\pm95$\\
Flaviviridae      &	Pestivirus     &	\verb;border          ;       &	--&	    12333&  $560\pm38$&  $681\pm10$&  $688\pm10$&        -- \\
Flaviviridae      &	Pestivirus     &	\verb;BVDV1           ;       &	--&	    12573&  $547\pm30$&  $692\pm10$&  $684\pm10$&        -- \\
Flaviviridae      &	Pestivirus     &	\verb;classicalSFV    ;       &	--&	    12301&  $617\pm61$&  $675\pm10$&  $667\pm98$&        -- \\
Flaviviridae      &	Pestivirus     &	\verb;pestiGiraffe    ;       &	--&	    12602&  $598\pm70$&  $693\pm11$&  $684\pm10$&        -- \\
Picornaviridae    &	Aphthovirus    &	\verb;BovRBV          ;       &	--&	     7556&  $375\pm36$&  $486\pm76$&  $474\pm77$&   $444\pm68$\\
Picornaviridae    &	Aphthovirus    &	\verb;ERAV            ;       &	2wff&	     7734&  $430\pm33$&  $508\pm81$&  $483\pm75$&   $518\pm82$\\
Picornaviridae    &	Aphthovirus    &	\verb;FMDV_typeO      ;       &	1zba&	     8134&  $406\pm31$&  $521\pm81$&  $517\pm79$&   $504\pm86$\\
Picornaviridae    &	Cardiovirus    &	\verb;saffold         ;       &	--&	     8115&  $487\pm36$&  $523\pm82$&  $504\pm78$&   $485\pm73$\\
Picornaviridae    &	Cardiovirus    &	\verb;TMEVlike        ;       &	--&	     7961&  $539\pm37$&  $513\pm82$&  $512\pm84$&   $494\pm76$\\
Picornaviridae    &	Enterovirus    &	\verb;BEV             ;       &	1bev&	     7414&  $462\pm47$&  $497\pm79$&  $484\pm76$&   $495\pm88$\\
Picornaviridae    &	Enterovirus    &	\verb;Hentero107      ;       &	--&	     7423&  $539\pm31$&  $487\pm77$&  $480\pm77$&   $474\pm71$\\
Picornaviridae    &	Enterovirus    &	\verb;Hrhino14        ;       &	1d3i&	     7212&  $419\pm17$&  $458\pm73$&  $449\pm71$&   $437\pm63$\\
Picornaviridae    &	Erbovirus      &	\verb;ERBV1           ;       &--&	     8828&  $393\pm27$&  $548\pm90$&  $549\pm86$&   $538\pm80$\\
Picornaviridae    &	Kobuvirus      &	\verb;aichi           ;       &	--&	     8251&  $235\pm20$&  $508\pm79$&  $491\pm78$&   $421\pm63$\\
Picornaviridae    &	Kobuvirus      &	\verb;bovineKV        ;       &	--&	     8374&  $405\pm28$&  $533\pm82$&  $539\pm79$&   $470\pm66$\\
Picornaviridae    &	Kobuvirus      &	\verb;porcineKV       ;       &	--&	     8210&  $266\pm25$&  $516\pm79$&  $499\pm80$&   $445\pm75$\\
Picornaviridae    &	Parechovirus   &	\verb;ljungan         ;       &	--&	     7590&  $425\pm36$&  $490\pm77$&  $473\pm75$&   $478\pm73$\\
Picornaviridae    &	Sapelovirus    &	\verb;asapelo         ;       &	--&	     8289&  $433\pm38$&  $520\pm82$&  $506\pm77$&   $506\pm75$\\
Picornaviridae    &	Senecavirus    &	\verb;SVV             ;       &	3cji&	     7310&  $364\pm24$&  $480\pm76$&  $475\pm76$&   $454\pm72$\\
Picornaviridae    &	Teschovirus    &	\verb;ptescho1        ;       &	--&	     7117&  $297\pm23$&  $482\pm78$&  $480\pm73$&   $498\pm84$\\
Picornaviridae    &	Tremovirus     &	\verb;AEV             ;       &	--&	     7055&  $425\pm43$&  $487\pm77$&  $469\pm74$&   $488\pm74$\\
Tombusviridae     &	Aureusvirus    &	\verb;MaWLMV          ;       &	--&	     4293&  $350\pm15$&  $357\pm56$&  $356\pm55$&	    --\\
Tombusviridae     &	Aureusvirus    &	\verb;pothos          ;       &	--&	     4354&  $358\pm18$&  $361\pm57$&  $358\pm56$&	    --\\
Tombusviridae     &	Avenavirus     &	\verb;OCSV            ;       &	--&	     4114&  $327\pm18$&  $331\pm50$&  $324\pm51$&	    --\\
Tombusviridae     &	Carmovirus     &	\verb;angelonia       ;       &	--&	     3964&  $338\pm16$&  $322\pm52$&  $319\pm49$&	    --\\
Tombusviridae     &	Carmovirus     &	\verb;JapINRV         ;       &	--&	     4014&  $326\pm45$&  $331\pm52$&  $327\pm55$&	    --\\
Tombusviridae     &	Carmovirus     &	\verb;PelFBV          ;       &	--&	     3923&  $266\pm13$&  $336\pm53$&  $327\pm52$&	    --\\
Tombusviridae     &	Carmovirus     &	\verb;TuCrV           ;       & 3zx8&	     4050&  $332\pm25$&  $333\pm54$&  $326\pm53$&	    --\\ %
Tombusviridae     &	Necrovirus     &	\verb;TNV_A           ;       &	1tnv&	     3684&  $269\pm6 $&  $298\pm47$&  $296\pm49$&	    --\\
Tombusviridae     &	Tombusvirus    &	\verb;GrALV           ;       &	--&	     4731&  $291\pm22$&  $375\pm59$&  $384\pm60$&	    --\\
Tombusviridae     &	Tombusvirus    &	\verb;pearLV          ;       &	--&	     4766&  $261\pm12$&  $367\pm59$&  $369\pm57$&	    --\\
Tymoviridae       &	Maculavirus    &	\verb;GFkV            ;       &	--&	     7564&  $250\pm20$&  $440\pm69$&  $457\pm69$&	    --\\
Tymoviridae       &	Marafivirus    &	\verb;GVSV1           ;       &	--&	     6506&  $392\pm36$&  $446\pm68$&  $437\pm67$&	    --\\
Tymoviridae       &	Marafivirus    &	\verb;MRFV            ;       &	--&	     6305&  $451\pm23$&  $443\pm67$&  $450\pm71$&	    --\\
Tymoviridae       &	Marafivirus    &	\verb;OBDV            ;       &	--&	     6509&  $328\pm35$&  $432\pm72$&  $424\pm61$&	    --\\
Tymoviridae       &	Marafivirus    &	\verb;OLV3            ;       &	--&	     7148&  $312\pm27$&  $429\pm67$&  $426\pm68$&	    --\\
Tymoviridae       &	Tymovirus      &	\verb;AnVYV           ;       &	--&	     6151&  $250\pm17$&  $356\pm56$&  $357\pm55$&	    --\\
Tymoviridae       &	Tymovirus      &	\verb;ChYMV           ;       &	--&	     6517&  $217\pm16$&  $357\pm58$&  $339\pm53$&	    --\\
Tymoviridae       &	Tymovirus      &	\verb;DiYMV           ;       &	--&	     6290&  $223\pm26$&  $361\pm56$&  $353\pm55$&	    --\\
Tymoviridae       &	Tymovirus      &	\verb;DuMV            ;       &	--&	     6181&  $336\pm44$&  $384\pm60$&  $384\pm59$&	    --\\
Tymoviridae       &	Tymovirus      &	\verb;EgMV            ;       &	--&	     6331&  $186\pm18$&  $352\pm57$&  $346\pm53$&	    --\\
Tymoviridae       &	Tymovirus      &	\verb;ErLV            ;       &	--&	     6035&  $248\pm24$&  $373\pm60$&  $368\pm59$&	    --\\
Tymoviridae       &	Tymovirus      &	\verb;NeRNV           ;       &	--&	     6285&  $302\pm23$&  $361\pm56$&  $351\pm53$&	    --\\
Tymoviridae       &	Tymovirus      &	\verb;OkMV            ;       &	--&	     6223&  $188\pm29$&  $333\pm52$&  $333\pm50$&	    --\\
Tymoviridae       &	Tymovirus      &	\verb;OnYMV           ;       &	--&	     6211&  $247\pm31$&  $384\pm62$&  $373\pm57$&	    --\\
Tymoviridae       &	Tymovirus      &	\verb;PlMV            ;       &	--&	     6154&  $369\pm26$&  $393\pm61$&  $389\pm64$&	    --\\
Tymoviridae       &	Tymovirus      &	\verb;ScMV            ;       &	--&	     6206&  $217\pm18$&  $348\pm54$&  $343\pm53$&	    --\\
\bottomrule\\
\caption{{\bf Set Viral genomes used in this study, including genome length
  and average MLD values.} $\langle
  MLD_{WT}\rangle $ refers to thermal average 
  of the MLD obtained on WT sequences.  $\overline{\langle MLD_{mut}\rangle}$, $\overline{\langle MLD_{UTRs}\rangle}$, 
  $\overline{\langle MLD_{CB}\rangle}$,  refer to average MLD values
  obtained on  synonymously
        mutated sequences, synonymously mutated sequences with preserved
        UTRs, and synonymously mutated sequences with preserved  UTRs and
        codon bias, respectively (see Material and Methods in the main text);
        in these cases an additional averaging over a wide set of
        possible mutations is performed. Errors are reported as standard
      deviations.}

\end{longtable}

%table of viruses used 